\documentclass{elsart}

\usepackage{harvard}

\usepackage{graphicx}

\usepackage{amssymb}

\def\url#1{{\ttfamily\def\/{/\discretionary{}{}{}}#1}}

\begin{document}

\begin{frontmatter}
\title{On the lives of extra-galactic radio sources:\\ the first 100,000 years}


\author[Snellen]{Ignas Snellen\thanksref{is}} and
\author[Schilizzi]{Richard Schilizzi}

\thanks[is]{This research was supported by the European Commission, 
TMR Programme, Research Network Contract ERBFMRXCT96-0034 ``CERES''}

\address[Snellen]{Institute of Astronomy, Madingley Road, Cambridge CB4 3EX, UK}
\address[Schilizzi]{Joint Institute for VLBI in Europe, Postbus 2, 7990 AA Dwingeloo, The Netherlands}

\begin{abstract}
In this paper we discuss the early phase of radio source evolution 
as represented by Gigahertz Peaked Spectrum (GPS) and Compact Steep Spectrum
(CSS) radio sources. Correlations between their spectral peak and angular size
strongly suggest that the spectral turnovers are caused by synchrotron self
absorption, and indicate that young radio sources evolve in a self similar 
way. 
We argue that the evolution of a radio source during its first $10^5$ years 
is  qualitatively very different from that during the rest of its life-time.
This may be caused by the difference in the density gradient of 
the intra-galactic medium inside and outside the core-radius of the 
host galaxy.
\end{abstract}

\end{frontmatter}

\section{Introduction}
\label{intro}
Gigahertz Peaked Spectrum (GPS) sources are characterised by 
a convex shaped radio spectrum peaking at about 1 GHz in frequency \cite{O1}. 
The existence of spectral turnovers in these objects implies that their radio 
emission is 
confined to very compact regions. Indeed, VLBI 
observations show that their radio structures are in general smaller than 
a few hundred parsecs \cite{St1}. The morphologies of GPS sources optically 
identified with galaxies are typically dominated by two components which
are more or less equal in flux density and spectral index. 
Since sometimes a very compact flat spectrum component can be seen
in the center, characteristic of a core, the two dominant outer components
 are in general interpreted as the hot-spots/mini-lobes.
The two-sided morphology of these objects is 
very distinctive compared to that of compact radio sources in general. 
When
selected from VLBI surveys, they are treated as 
a separate class \cite{W1}, and named Compact Symmetric Objects (CSO).
The overlap between CSO and GPS galaxies is 
large and it can be assumed they form
one and the same class of object. However, the overlap is not complete:
orientation effects can alter both the morphology and radio spectrum in such
way that the objects are not classified as CSO or GPS respectively \cite{S1}.
Furthermore, it has been claimed that some CSO do not exhibit a spectral 
turnover at low frequencies. If true, this would make the young nature
of these particular CSOs less likely, since it implies the existence of 
substantial large scale radio emission. GPS sources identified with
quasars, which are mostly found at high redshifts, have 
core-jet morphologies in general and may not be physically related to the 
CSO/GPS galaxies\cite{St1,S2}.

Since the initial discovery of Gigahertz Peaked Spectrum (GPS) sources, 
it has been speculated that these are young objects \cite{SH,BL}, but
only recently, compelling evidence in favour of this hypothesis has been 
given. The strongest evidence comes from measurements of the 
separation velocities of the hot-spots in several GPS sources, implying
dynamic ages of typically $10^3$ years (see Conway, this volume).
Furthermore, detailed measurements of the spectral ages of the somewhat 
larger Compact Steep Spectrum (CSS) sources indicate spectral
ages in the range of $10^3-10^5$ years (see Murgia, this volume).
The alternative hypothesis that GPS and CSS sources are old
objects situated in a very dense environment impeding the outward 
motion of the jet is also less likely, since no evidence for 
any difference between the environments of GPS, CSS and large size 
radio sources has been found.

In this paper, we discuss results on the early evolution of radio sources
from the investigation of three samples of faint and bright GPS and CSS 
galaxies:
1) The faint GPS sample selected from the WENSS survey \cite{S2,S3,S4},
2) the bright GPS sample from Stanghellini et al. (1998), and 3) the CSS
sample selected by Fanti et al. (1990).

\section{The spectral turnovers and morphological evolution}

\begin{figure}
\begin{center}
\includegraphics*[width=9cm,angle=-90]{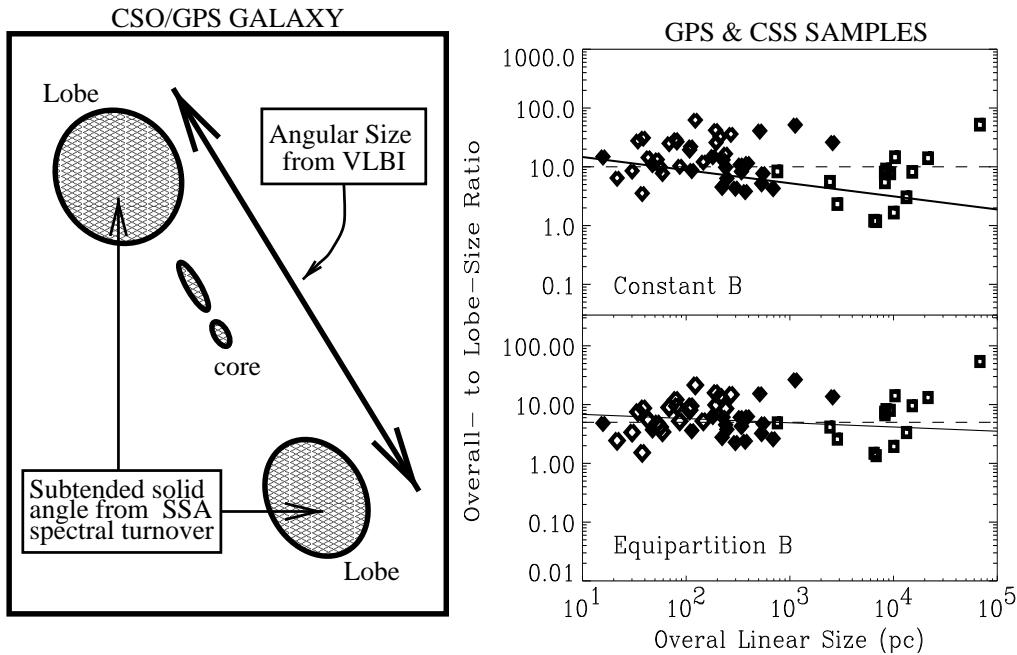}
\end{center}
\caption{(Left) A schematic view of a young radio source. The overall angular 
size can be determined from VLBI observations, while the sizes of the lobes 
can be derived from the synchrotron self-absorption turnover. 
(Right) The overall size to lobe size ratio in faint and bright samples of 
GPS and CSS galaxies, assuming a fixed- (top panel) and 
an equipartition magnetic field (bottom panel).}
\label{fig1}
\end{figure}

The combination of the faint GPS sample and bright GPS and CSS samples from
the literature gave a unique opportunity to investigate the relation 
between the spectral peak and size of young radio sources.
Not surprisingly, the inverse correlation between peak frequency, $\nu_p$, and
 angular size, $\theta$, was confirmed. However, in addition, a correlation 
was found between the peak flux density, $S_p$, and angular size. 
The strengths and signs of these two correlations are exactly as expected for 
synchrotron self absorption (SSA), for which $\theta^2 \propto S_p B^{1/2} 
\nu_p^{-5/2}$, where $B$ is the magnetic field. This strongly suggests that 
SSA is indeed the cause of the spectral turnovers in GPS and CSS sources, and
not free-free absorption as recently proposed by Bicknell et al. (1997).

The solid angle subtended by the dominant features, the mini-lobes, 
determines the strength and frequency of the spectral peak 
(see fig \ref{fig1}). The angular size determined from VLBI observations, 
corresponds to the {\it overall} angular size of the object. 
The correlations discussed above therefore imply a constant ratio
of overall size to lobe size in samples of faint and bright GPS and CSS
galaxies. This indicates that young radio sources grow in a self-similar 
way. 

The sizes derived for the lobes are slightly dependent on 
the magnetic field strength (see above). The ratios of the overall
to lobe sizes were determined for a fixed 
magnetic field of $10^{-3}$ Gauss and for an equipartition magnetic field 
\cite{SC}. The results are shown in the top and bottom panel
of fig \ref{fig1} respectively. The ratios, calculated using an equipartition
magnetic field, do not show a trend with linear size, but
they show a decline with linear 
size when a fixed magnetic field is used. 
This seems to indicate that the self-similar evolution scenario 
(dashed line) is better fitted (solid line) for an equipartition 
magnetic field than for a constant magnetic field.

\section{The Luminosity evolution of radio galaxies}

\begin{figure}
\begin{center}
\includegraphics*[width=14cm]{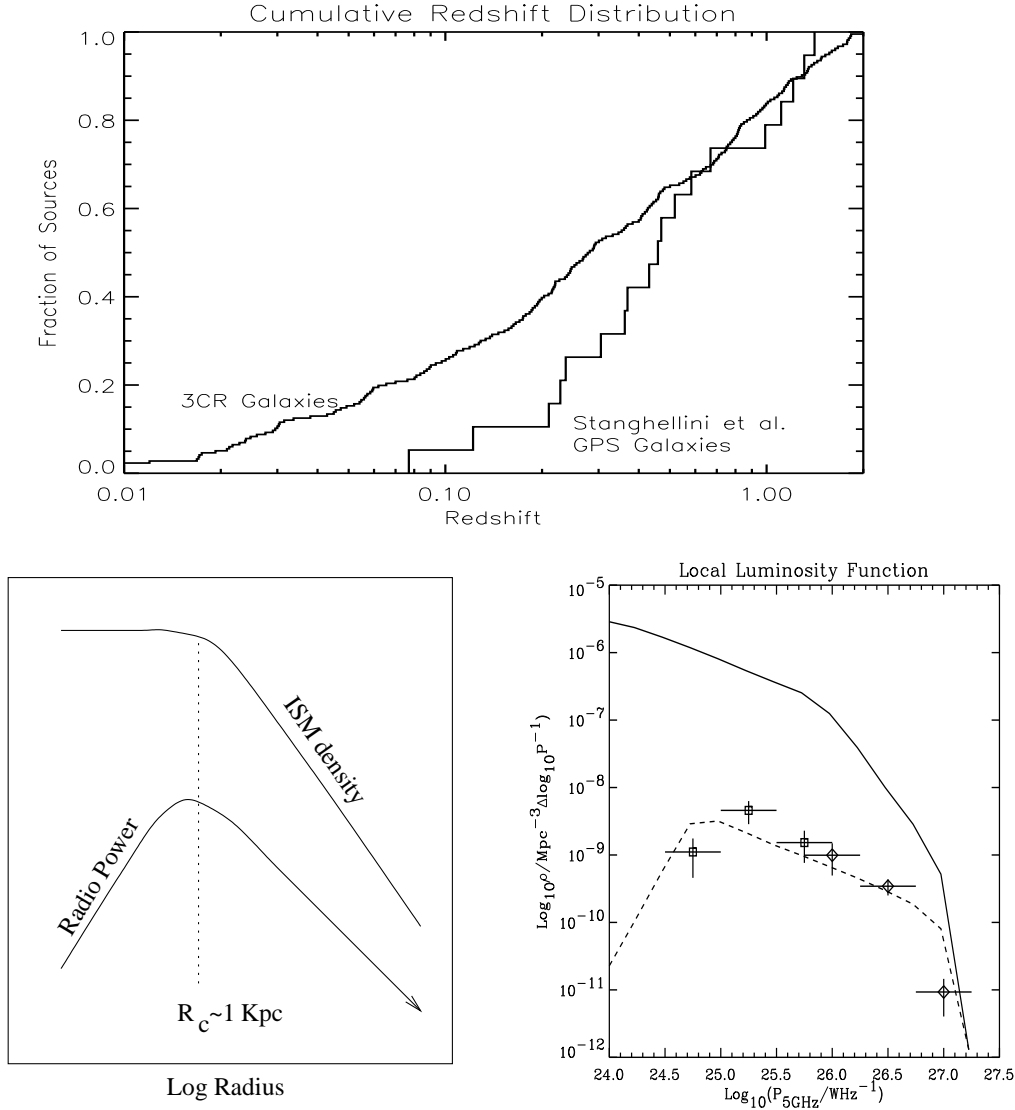}
\end{center}
\caption{(Top) The cumulative redshift distributions of GPS galaxies from
the Stanghellini et al. sample (1998) and of 3CR galaxies. (Left) Schematic
representation of the intra-galactic medium and the proposed luminosity
evolution scenario for radio sources. (Right) The Local Luminosity Function
of young radio sources derived from the faint and bright GPS samples. The solid
and dashed lines represent simulated luminosity functions for extended and young objects respectively.}
\label{fig2}
\end{figure}

Several evolution models have been proposed for GPS sources, 
in which GPS sources subsequently evolve into CSS sources and 
large-scale doubles \cite{HM,F2,R1,OB}. 
In these models, the age ratio of large size objects to GPS sources 
is typically $\sim 10^3$. The much larger fraction (say 10\%) 
of GPS in radio surveys therefore implies that young radio
sources have to substantially decrease in radio luminosity (a factor $\sim 10$)
when evolving to large scale radio sources.
Readhead et al. (1996) find from their CSO statistics that the luminosity
evolution from 10 pc to 150 Kpc is consistent with a single power-law 
decrease. This in contrast to O'Dea \& Baum (1997) who find that the number of 
GPS and CSS sources per bin of log projected size is constant from 100 pc to
6 Kpc, indicating that GPS and CSS sources must decline in luminosity at a
faster rate than the classical 3CR doubles. 
The number count and linear size statistics used in these studies,  
are all averaged over a wide redshift range and only cover 
the brightest objects in the sky. 
However, as is shown in figure \ref{fig2}, in flux density limited samples 
the redshift 
distribution of GPS galaxies is significantly different from that of large 
size radio galaxies. This suggests that the interpretation of the number 
count statistics is not so straightforward.

The bias of GPS galaxies towards higher redshifts than large size radio 
galaxies provides an important clue about the luminosity evolution of 
radio sources. It implies that GPS galaxies are biased towards higher
radio power than extended objects in flux density limited
samples. If GPS and large size radio sources are identical objects, just 
observed at different phases pf the life cycle, their cosmological density 
evolution, e.g. 
their birth functions with redshift, should be the same. Since their 
lifetimes are short compared to the Hubble time, the redshift distributions 
of the GPS galaxies, and the objects they evolve to, should 
also be the same. The bias of GPS sources towards higher redshifts and 
radio powers 
therefore implies that their luminosity function must be flatter than
that of large size radio sources. We argue that the luminosity evolution
of the individual objects strongly influences their collective luminosity 
function, and propose an evolution scenario in which GPS sources 
increase in luminosity and large size sources decrease in luminosity with time.
In the simplified case, in which source to source variations in
the surrounding medium can be ignored,
the luminosity of a radio source only depends on its age and jet power.
Sources in a volume based sample are biased towards older ages and lower 
jet powers for populations of both GPS and large size sources.
Low jet powers result in low luminosity sources. The higher the age of a 
large size source, the lower its luminosity, but the higher the age of a
GPS source, the higher its luminosity. This means that for a population of 
large size sources the jet power and age bias strengthen each other resulting
in a steep luminosity function, while they counteract for GPS sources,
resulting in a flatter luminosity function.

The luminosity evolution proposed is expected for a ram-pressure confined
radio source in a surrounding medium with a King-profile density. 
In the inner parts of the King profile, the density of the medium is 
constant and the radio source builds up its luminosity, but after it grows
large enough the density of the medium declines and the luminosity of the radio
source decreases. An analytic model for radio sources with pressure confined
jets has been developed by Kaiser \& Alexander (1997). Interestingly, they showed 
that the properties of the bow shock and the surrounding gas force radio sources
to grow in a self-similar way, provided that the density of the surrounding gas
falls off less steeply than $1/r^2$. X-ray observations of large nearby ellipticals
show that their hot ISM follows a King distribution well, and have a core 
radius of typically 500-1000 pc \cite{t1}. Hence the 
predicted change in luminosity evolution can be expected to occur between 
the GPS and the CSS phases.

A way to test this luminosity scenario is to determine
the local luminosity functions (LLF) for GPS galaxies and large size radio 
sources and compare it with simulated luminosity functions for a population
of radio sources undergoing the proposed luminosity evolution.
Unfortunately, only 4 GPS galaxies at $z<0.2$ are present in the combined
faint and bright GPS samples, too small a number to construct a LLF directly.
However, since it can be assumed that the cosmological number density evolution
for the young sources is the same as for old sources, the cosmological 
evolution of the luminosity function as derived for steep spectrum (eg. large
size) sources \cite{D1} can be used to derive a LLF for
young radio sources from the total faint and bright GPS samples.
The combination of the bright and faint GPS samples is not straightforward,
since they are selected in very different ways. This introduces a
relatively uncertain correction factor of $\sim3$ for the number densities
derived from the faint sample. The result is shown in figure \ref{fig2}.
A radio source population is simulated 
having random ages between 0 and 1000 time-units and jet powers
over a range of 200, distributed with a powerlaw of -1.69, chosen to 
result in a slope of the luminosity function of large size sources at low 
radio powers of 0.69 in log, as determined  by Dunlop \& Peacock (1990).
 The objects younger than 1 time-unit are designated as GPS sources, and 
increase 
in luminosity with time, while the older sources decrease in radio luminosity.
Assuming that the boundary between the GPS and large size phases 
is at $10^5$ year, the age-limit of the large size radio sources in the 
simulation
is $10^8$ years. The resulting luminosity functions were scaled
in such way that the break in the luminosity function of the large scale 
sources overlaps with what is found by  Dunlop \& Peacock for steep spectrum 
sources.
Although the uncertainties are large and several free
parameters enter the simulation, figure \ref{fig2} shows that the shape of the 
LLF of GPS galaxies is as expected. 
This scenario is also consistent with the high number densities of 
GPS sources at bright flux density levels, since at the high luminosity end,
the simulated LLF of GPS galaxies is only slightly lower than that of large 
size galaxies.

\end{document}